\documentclass[appendixfloats]{emulateapj}
\usepackage[titletoc,title]{appendix}
\usepackage{longtable}
\usepackage{ltxtable} 
\usepackage{graphicx}
\usepackage{graphics}
\usepackage{color}
\usepackage{epsfig}
\usepackage{rotating}
\usepackage{subfigure}
\usepackage{float}
\usepackage{url}
\usepackage{hyperref}
\usepackage{breakurl}
\usepackage{scalerel}

\shortauthors{J. T. Li et al.}
\shorttitle{CGM-MASS III: Stacking Analaysis}

\begin{document}

\title{Baryon Budget of the Hot Circumgalactic Medium of Massive Spiral Galaxies}

\author{Jiang-Tao Li\altaffilmark{1}, Joel N. Bregman\altaffilmark{1}, Q. Daniel Wang\altaffilmark{2}, Robert A. Crain\altaffilmark{3}, and Michael E. Anderson\altaffilmark{4}} 

\altaffiltext{1}{Department of Astronomy, University of Michigan, 311 West Hall, 1085 S. University Ave, Ann Arbor, MI, 48109-1107, U.S.A.}

\altaffiltext{2}{Department of Astronomy, University of Massachusetts, 710 North Pleasant St., Amherst, MA, 01003, U.S.A.}

\altaffiltext{3}{Astrophysics Research Institute, Liverpool John Moores University, IC2, Liverpool Science Park, 146 Brownlow Hill, Liverpool, L3 5RF, United Kingdom}

\altaffiltext{4}{Max-Planck Institute for Astrophysics, Karl-Schwarzschild-Stra$\rm\beta$e 1, 85748 Garching bei M\"{u}nchen, Germany}

\keywords{\emph{(galaxies:)} intergalactic medium --- X-rays: galaxies --- galaxies: haloes --- galaxies: spiral --- galaxies: evolution --- galaxies: fundamental parameters.}

\nonumber

\begin{abstract}
The baryon content around local galaxies is observed to be much less than is needed in Big Bang nucleosynthesis. Simulations indicate that a significant fraction of these ``missing baryons'' may be stored in a hot tenuous circum-galactic medium (CGM) around massive galaxies extending to or even beyond the virial radius of their dark matter halos. Previous observations in X-ray and Sunyaev-Zel'dovich (SZ) signal claimed that $\sim(1-50)\%$ of the expected baryons are stored in a hot CGM within the virial radius. The large scatter is mainly caused by the very uncertain extrapolation of the hot gas density profile based on the detection in a small radial range (typically within 10\%-20\% of the virial radius). Here we report stacking X-ray observations of six local isolated massive spiral galaxies from the CGM-MASS sample. We find that the mean density profile can be characterized by a single power law out to a galactocentric radius of $\approx 200\rm~kpc$ (or $\approx130\rm~kpc$ above the 1~$\sigma$ background uncertainty), about half the virial radius of the dark matter halo. We can now estimate that the hot CGM within the virial radius accounts for $(8\pm4)\%$ of the baryonic mass expected for the halos. Including the stars, the baryon fraction is $(27\pm16)\%$, or $(39\pm20)\%$ by assuming a flattened density profile at $r\gtrsim130\rm~kpc$. We conclude that the hot baryons within the virial radius of massive galaxy halos are insufficient to explain the ``missing baryons''. 
\end{abstract}

\section{Introduction}\label{sec:Introduction}

Multi-wavelength observations have been conducted to measure the mass of baryons in different forms such as stars and different phases of the interstellar, circum-galactic, and intra-cluster medium (ISM, CGM, and ICM; e.g., \citealt{Bregman07,Anderson10,Chiu17}). Combining these phases, the best observed case, our Milky Way (MW), has $<50\%$ of the expected baryons detected (e.g., \citealt{Miller13,Miller15}). The ``missing baryon'' problem is more severe for less massive galaxies. The undetected baryons are expected to be distributed in larger scale structures (e.g., \citealt{Haider16,deGraaff17}) or in a less readily-detected phase such as clouds of warm gas ($10^{4-5.5}\rm~K$, e.g., \citealt{Tumlinson11,Lim17b}) or a dilute hydrostatic halo of hot gas near the virial temperature ($10^{5.5-6.8}\rm~K$; e.g., \citealt{Crain07,Faerman17}). The underlying physics regulating the baryon budget becomes the most uncertain ingredients of the current galaxy formation models.

Recent years have borne witness to significant progress in searches for the missing baryons, via ultraviolet (UV) absorption line measurements of the cool CGM ($<10^5\rm~K$) and the warm-hot intergalactic medium (WHIM), and microwave measurements of the SZ signals and X-rays from the hot CGM. But different observations show large scatters in the measured baryon budget, and even analysis of the same data by different groups produces significantly different results (e.g., \citealt{Anderson13,Planck13,Werk14,Keeney17,Lim17a}). In particular, the inferred hot gas mass depends critically on the very uncertain extrapolation from the observed region typically at $r\lesssim(0.1-0.2)r_{\rm 200}$ ($r_{\rm 200}$ approximately equals to the virial radius of the dark matter halo) to $r\approx r_{\rm 200}$, and could vary by at least a factor of a few based on different assumptions for the radial density profile (e.g., \citealt{Dai12,Faerman17}).

Massive isolated spiral galaxies provide the best cases to search for the missing baryons in the extended hot CGM. They are massive enough to gravitationally heat the infalling gas to X-ray emitting temperatures and/or to confine the volume-filling gas with $T\approx {\rm a~few}\times10^6\rm~K$. Their star formation is often largely quenched so they have little contamination from the metal-enriched stellar feedback material to their halos, which often dominates the X-ray emission despite its relatively low mass \citep{Crain13}. Compared to massive elliptical galaxies, they also exhibit relatively simple formation histories without recent major mergers, and occupy low density environments with little contamination from the ICM. Efforts have been made to study the hot gas associated with massive isolated spirals, which to date have resulted in the detection of extended emission to $r=(50-70)\rm~kpc$ in a few cases (e.g., \citealt{Dai12,Bogdan13,Bogdan17,Anderson16}).

We studied the hot gaseous halo of six isolated spiral galaxies, which are among the most massive in the local universe ($d<100\rm~Mpc$), with a stellar mass $M_*\gtrsim1.5\times10^{11}\rm~M_\odot$ and rotation velocity $v_{\rm rot}>300\rm~km~s^{-1}$. This study is based on the analysis of the observations from a large \emph{XMM-Newton} program conducted in AO-13/14 [five galaxies; The \textbf{C}ircum-\textbf{G}alactic \textbf{M}edium of \textbf{MAS}sive \textbf{S}pirals (CGM-MASS) sample; \citealt{Li16b,Li17}] and from the \emph{XMM-Newton} archive for one galaxy \citep{Dai12}. These observations detect X-ray emission from the hot gas out to $r\approx 50\rm~kpc$ around individual galaxies \citep{Li16b,Li17}. Here we report the outcome of stacking all six CGM-MASS galaxies to achieve an unprecedented sensitivity for detecting the low-surface brightness emission of the hot CGM. The errors quoted in this letter are at $1~\sigma$ confidence level and are statistical only. Systematical uncertainties such as the intrinsic uncertainties of the stellar and background models are discussed in \citet{Li16b,Li17} and are in general not large enough to affect the detection of the large scale features. We will also discuss other systematical uncertainties related to the stacking in \S\ref{sec:DataReduction}, and present the results on the hot baryon budget in \S\ref{sec:Discussion}.

\section{Data reduction and systematic errors}\label{sec:DataReduction}

We present a careful stacking analysis of the CGM-MASS sample with a total effective \emph{XMM-Newton} exposure of $\sim0.6\rm~Ms$. Details of data reduction on individual galaxies, as well as statistical analysis comparing the CGM-MASS sample to other samples, are presented in \citet{Li16b,Li17}. In particular, the radial X-ray intensity profiles are extracted from all of the three EPIC instruments (MOS-1, MOS-2, and PN) and stacked based on the same distance scale. We do not account for the azimuthal variations of the X-ray intensity, which is not significant even on galaxy disk scales in these quiescent galaxies \citep{Li16b}. Comparing to stacking analysis of the survey data such as from the \emph{ROSAT} \citep{Anderson13}, we benefit from cleanly removing all the resolved background sources before stacking. The stacking is done for two slightly different radial scaling schemes: one is scaled to the same physical distance (in kpc) and the other to the same halo mass-dependent virial radius (in $r_{\rm 200}$). 

\begin{figure*}
\begin{center}
\epsfig{figure=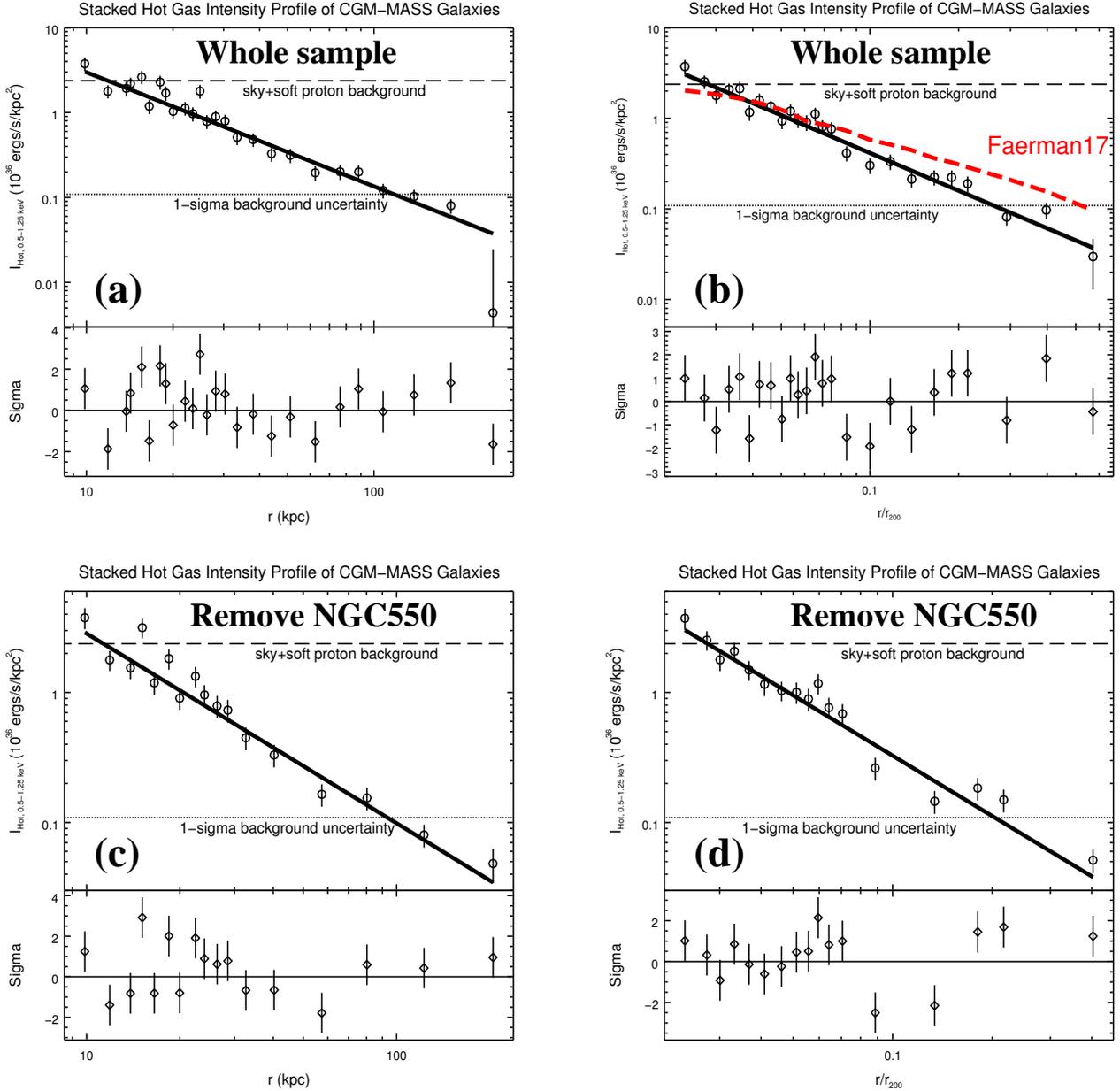,width=1.0\textwidth,angle=0, clip=}
\caption{Stacked radial intensity profiles of the hot gas component of the CGM-MASS galaxies. Different panels have the galactocentric radial distance of different galaxies rescaled to kpc (a,c) and $r_{\rm 200}$ (b,d). The solid line is the best-fit $\beta$-function. The dashed and dotted lines show the sky+soft proton background and the 1~$\sigma$ uncertainty. Errors are statistical only. In panel (b), the red dashed curve is the \citet{Faerman17} model scaled with $r_{\rm 200}$ of a MW-sized halo and renormalized to fit the data at $r<0.08r_{\rm200}\approx20\rm~kpc$. Such a model has significantly flatter X-ray intensity profile at large radii than what is measured for the more massive CGM-MASS galaxies, but such a difference is not visible at $r\lesssim0.1r_{\rm200}$. In panels (c) and (d), NGC~550 has been removed from the stacking because there is an X-ray luminous background cluster Abell~189 projected close to this galaxy \citep{Li17}.
}\label{fig:rProfileStack}
\end{center}
\end{figure*}

As shown in Fig.~\ref{fig:rProfileStack}, the 0.5-1.25~keV X-ray intensity profile tracing the hot gas emission (energy band selected to avoid strong instrumental lines, e.g., \citealt{Li16b}) can be fitted with a $\beta$-function $I_{\rm X}=I_{\rm X,0}[1+(r/r_{\rm core})^2]^{0.5-3\beta}$ with $r_{\rm core}$ fixed at 1.0~kpc ($0.002r_{\rm 200}$); results are insensitive to the choice of $r_{\rm core}$. The profile follows the same slope to $r\sim200\rm~kpc$ or $r\sim0.5r_{\rm 200}$ (above the 1~$\sigma$ background uncertainty at $r\sim130\rm~kpc$ or $r\sim0.25r_{\rm 200}$), which more than doubles the radial range over which the emission around individual galaxies has been detected \citep{Li16b,Li17}. The best-fit slopes of the two profiles are indistinguishable: $\beta=0.391\pm{0.009}$ (Fig.~\ref{fig:rProfileStack}a) and $\beta=0.397\pm0.009$ (Fig.~\ref{fig:rProfileStack}b). The constant slope is clearly different from some models with a flattened density profile (e.g., \citealt{Faerman17}; Fig.~\ref{fig:rProfileStack}b).

We next discuss some systematical uncertainties related to the stacking analysis.

\subsection{Physical and chemical properties of hot gas}\label{subsection:SpecAnalysis}

\begin{figure*}
\begin{center}
\epsfig{figure=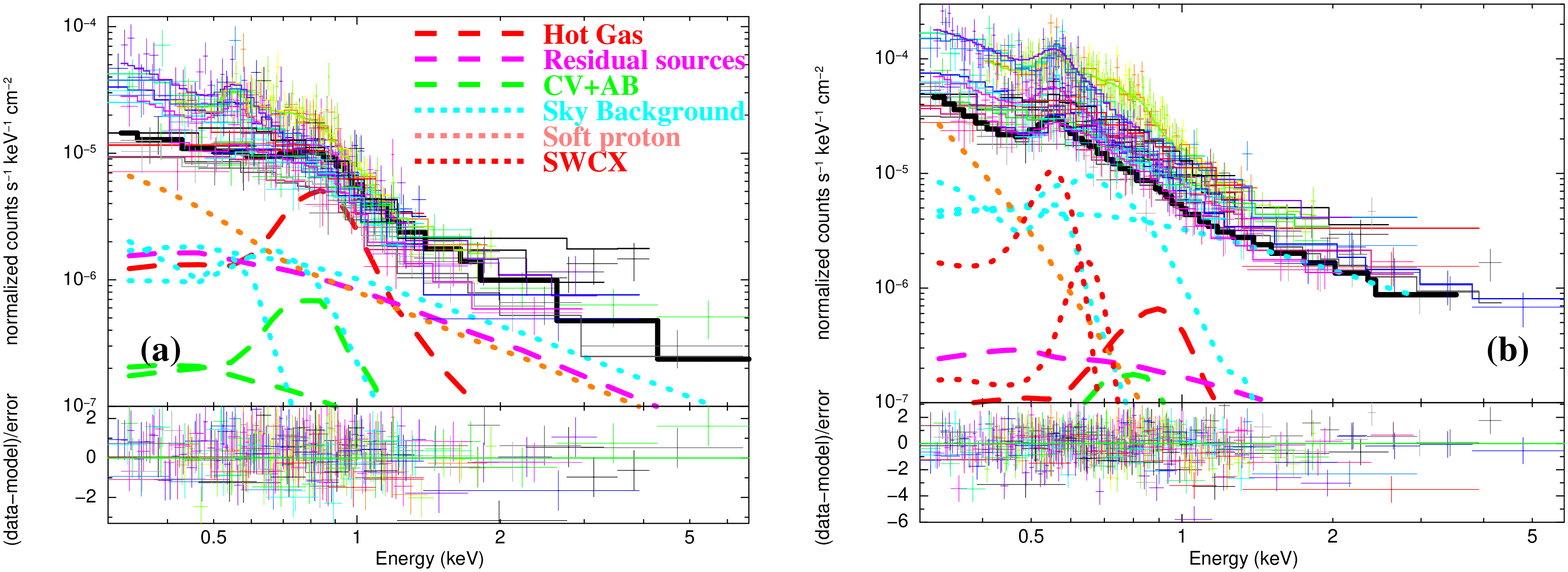,width=1.0\textwidth,angle=0, clip=}
\caption{Joint spectral analysis of the CGM-MASS galaxies. (a) and (b) are extracted from circular regions with $r<0.05r_{\rm 200}$ and $r=(0.05-0.1)r_{\rm 200}$, respectively. Data points with different colors represent spectra of different instruments (MOS-1, MOS-2, and PN) and different galaxies. The thick colored curves are different model components of the thick black curve representing the combined model of the PN/MOS-1 spectrum of UGC~12591/NGC~550 (the observation 0741300501) in panel (a)/(b). The colored curves are denoted on top right of (a). In particular, dotted curves represent different background components, including the sky background (MW halo, local hot bubble, distant AGN), the soft proton background, and the solar wind charge exchange (SWCX) background only present in the data of a few galaxies. Dashed curves are different source components, including the contributions from cataclysmic variables (CVs) and coronal active binaries (ABs) scaled from the K-band luminosity, the residual after removing bright point-like sources (described with a power law), and the hot gas emission. Details of spectral modeling are discussed in \citet{Li16b,Li17}.}\label{fig:CGMMASSSpec}
\end{center}
\end{figure*}

Physical and chemical properties of hot gas can be obtained by jointly analyzing the spectra of individual galaxies extracted from $r<0.05r_{\rm 200}$ (Fig.~\ref{fig:CGMMASSSpec}a) or $r=(0.05-0.1)r_{\rm 200}$ (Fig.~\ref{fig:CGMMASSSpec}b). Various source and background components and the spectral model describing them are discussed in \citet{Li16b,Li17}. Model parameters of various stellar sources and background components are all fixed and scaled for each galaxy, while temperature and metallicity of the hot gas (APEC model) are linked for different galaxies. These parameters, together with the emission measure of each galaxies, are the only free parameters in the joint spectral analysis. 

The best-fit hot gas temperature at $r<0.05r_{\rm 200}$ and $r=(0.05-0.1)r_{\rm 200}$ are $0.77\pm0.04\rm~keV$ and $0.87_{-0.13}^{+0.43}\rm~keV$, respectively. We therefore do not find any temperature gradient within $1~\sigma$ for the hot CGM around the CGM-MASS galaxies. The metallicity of hot gas ($Z_{\rm gas}$) is poorly constrained. We fix it at $Z_{\rm gas}=0.2Z_\odot$ throughout the halo when calculating other hot gas parameters. This assumption is based on estimates from X-ray observations of similar galaxies (e.g., \citealt{Bogdan13,Anderson16}) and is consistent with the apparently featureless X-ray spectra (strongly metal-enriched hot gas produces an easily detectable spectral feature at $\sim1\rm~keV$; e.g., \citealt{Li09,Li15a}).

\subsection{Scatter of hot gas properties of different galaxies}\label{subsection:SpatialAnalysis}

All the CGM-MASS galaxies are isolated spiral galaxies with similar stellar mass, SFR, and gravitational potential, so are expected to have similar hot gas properties (e.g., \citealt{Li13a,Li13b,Wang16}), which is not inconsistent with spectral analysis of individual galaxies, e.g., typically within $\sim2\rm~\sigma$ for the temperature \citep{Li17}. We therefore believe the link of the hot gas properties of different galaxies in spectral analysis does not bias the results. Such a joint spectral analysis is similar to stacking the X-ray spectra of a galaxy sample \citep{Anderson13}, but the CGM-MASS galaxies have less contamination from the metal-enriched starburst feedback material and the ICM. They also have more uniform galaxy properties than previous studies.

\begin{figure*}
\begin{center}
\epsfig{figure=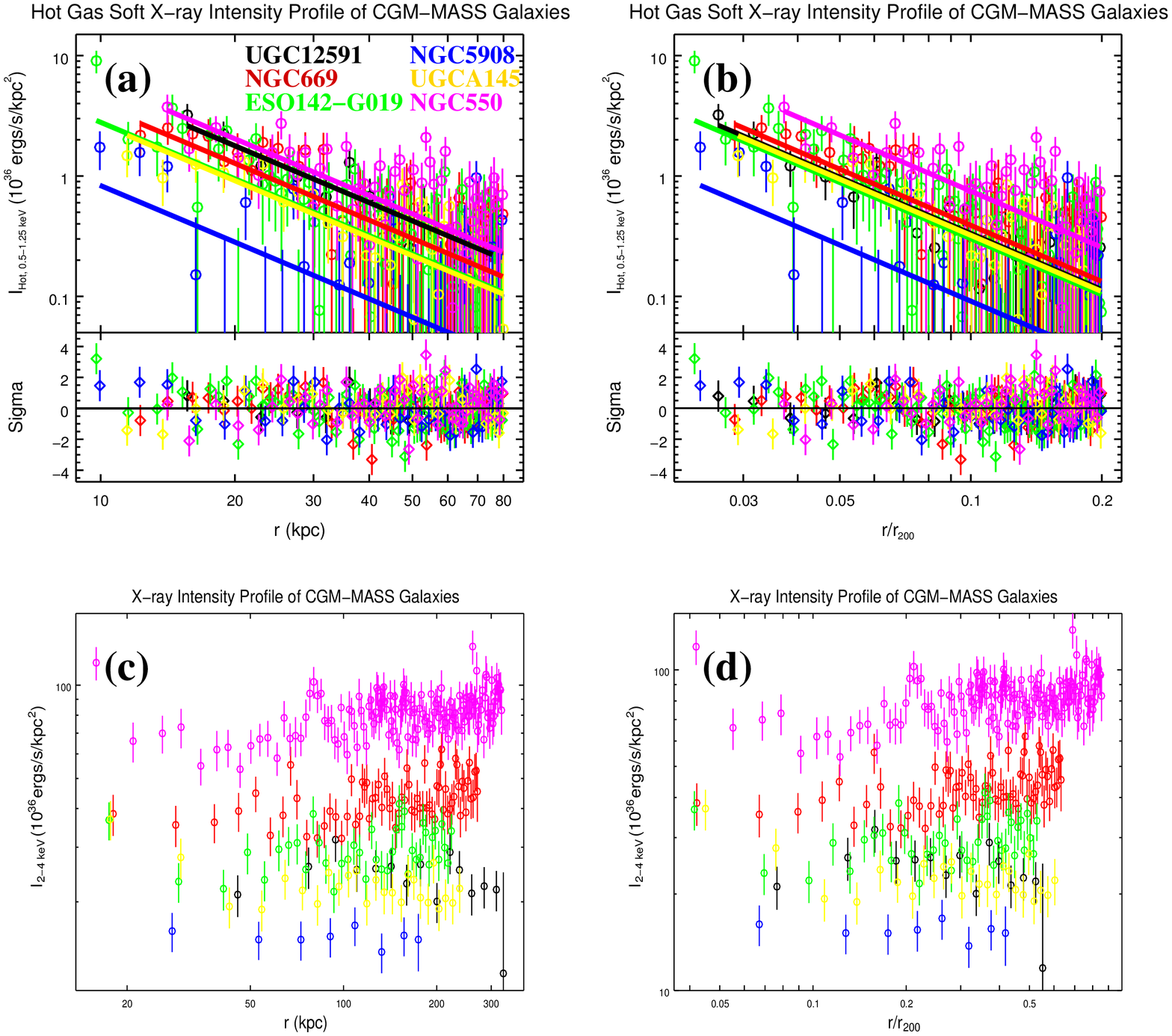,width=1.0\textwidth,angle=0, clip=}
\caption{X-ray intensity profiles of the CGM-MASS galaxies. Different colors represent data from different galaxies as denoted in (a). The galactocentric radius has been scaled in physical unit (kpc) in (a,c) while in $r_{\rm 200}$ in (b,d). For panels (a) and (b), the intensity is measured in 0.5-1.25~keV for the hot gas component only, after subtracting various stellar and background components. The solid lines are the best-fit models of each galaxies, with linked slope ($\beta$ index) but different normalizations. For panels (c) and (d), the intensity is measured in 2-4~keV. Stellar and sky+soft proton background components are not subtracted, but quiescent instrumental background is subtracted and standard exposure correction is adopted.}\label{fig:rProfileJointfit}
\end{center}
\end{figure*}

We further show the joint fit of the 0.5-1.25~keV intensity profile of the CGM-MASS galaxies after subtracting the stellar and background components (Fig.~\ref{fig:rProfileJointfit}). The slopes of the $\beta$-model fitting different galaxies are consistent with each other and can be linked. The consistency of the slope in different galaxies indicates that the stacking of the profiles in Fig.~\ref{fig:rProfileStack} does not bias the estimate of $\beta$. The best-fit $\beta$ is $0.428\pm{0.015}$ for Fig.~\ref{fig:rProfileJointfit}a and $0.425_{-0.014}^{+0.015}$ for Fig.~\ref{fig:rProfileJointfit}b, consistent with other massive spirals (e.g., \citealt{Dai12,Bogdan13}) and with the value we obtained in the stacking analysis (at $\lesssim2~\sigma$).

\subsection{Background sources}\label{subsection:removeN550}

There is an X-ray luminous background galaxy cluster, Abell~189, projected close to NGC~550 \citep{Li17}. Although we have masked the extended X-ray emission from this cluster, the low surface brightness residual emission may still affect our analysis of the X-ray intensity profile at large radii. We therefore removed NGC~550 and stacked the radial intensity profiles of the other five CGM-MASS galaxies in Fig.~\ref{fig:rProfileStack}c,d. The best-fit $\beta$ is $0.410_{-0.011}^{+0.012}$ for Fig.~\ref{fig:rProfileStack}c and $0.424_{-0.012}^{+0.013}$ for Fig.~\ref{fig:rProfileStack}d, consistent with the results including this galaxy within $\sim(1-2)~\sigma$. This examination also confirms that NGC~550, which has about twice the exposure time of other galaxies \citep{Li16b}, does not dominate the signal at large radii.

\subsection{Vignetting effect}\label{subsection:vignetting}

We generate exposure maps with standard \emph{XMM-Newton} data reduction software to correct the vignetting effect \citep{Li16b}. Because all the sample galaxies are located at the center of the \emph{XMM-Newton} field of view, the stacking of them may magnify the vignetting effect and may affect the observed radial distribution of hot gas. In principle, the vignetting effect is stronger at higher energies. We then examine how strong the residual vignetting effect may be by comparing the radial distribution of soft (0.5-1.25~keV, Fig.~\ref{fig:rProfileJointfit}a,b) and hard X-ray emissions (2-4~keV, Fig.~\ref{fig:rProfileJointfit}c,d). Because it is more difficult to determine the sky background in hard X-ray (due to stronger vignetting and soft proton emission), we do not subtract the sky background in Fig.~\ref{fig:rProfileJointfit}c,d. 

For all the CGM-MASS galaxies, the 2-4~keV intensity is roughly constant or even increases in some cases toward larger radii at $r\gtrsim50\rm~kpc$ or $r\gtrsim0.1r_{\rm 200}$ (Fig.~\ref{fig:rProfileJointfit}c,d). This is significantly different from the declining intensity to a much larger radius in soft X-ray (Fig.~\ref{fig:rProfileJointfit}a,b). Therefore, no matter what effects (such as vignetting) produce the shape of the hard X-ray intensity profile, similar effects cannot produce the declining soft X-ray intensity profile, which is assumed to be mainly a result of the declining hot gas density. 

\section{Discussion and Conclusions}\label{sec:Discussion}

In order to estimate the characteristic baryon budget of the CGM-MASS sample, we construct a fiducial galaxy that has their average properties ($M_*$, $v_{\rm rot}$, $r_{\rm 200}$, and $M_{\rm 200}$). We adopt an APEC model subjected to foreground extinction with $N_{\rm H}=5\times10^{20}\rm~cm^{-2}$, $kT=0.8\rm~keV$, and $Z_{\rm gas}=0.2Z_\odot$ to convert the soft X-ray intensity to an electron density. Based on the stacked X-ray intensity profile (adopting Fig.~\ref{fig:rProfileStack}b), we derive the radial distribution of various hot gas properties in the same fashion as for individual galaxies \citep{Li17} and summarize the integrated properties of the CGM of the fiducial galaxy in Table~\ref{table:FiducialGalaxy}. The errors of the parameters in Table~\ref{table:FiducialGalaxy} already include the uncertainty on $\beta$, but do not include the systematical uncertainties discussed in \S\ref{sec:DataReduction}, which are difficult to quantify. As an example, we show here how the uncertainty on the assumed gas metallicity $Z_{\rm gas}$ affects the estimated hot baryon mass. $Z_{\rm gas}$ anti-correlates with the gas density so the derived hot baryon mass. If $Z_{\rm gas}=0.1Z_\odot$ ($Z_{\rm gas}=Z_\odot$), the hot baryon mass will be $\sim$140\% (45\%) of the value obtained by assuming $Z_{\rm gas}=0.2Z_\odot$. This uncertainty, however, does not significantly affect our conclusions.

\begin{table}
\begin{center}
\caption{Properties of the Fiducial Galaxy} 
\footnotesize
\vspace{-0.0in}
\tabcolsep=3.2pt%
\begin{tabular}{lccccccccccccc}
\hline
Parameter        & Value \\
\hline
$\log M_*/M_\odot$    & $11.45\pm0.20$ \\
$v_{\rm rot}/{\rm km~s^{-1}}$   & $360.37\pm54.12$ \\
$\log M_{\rm 200}/M_\odot$     & $12.96\pm0.21$ \\
$r_{\rm 200}/\rm kpc$     & $433.32\pm70.06$ \\
$kT/\rm keV$ & 0.8 \\
$Z/Z_\odot$ & 0.2 \\
$I_{\rm 0}/(10^{38}\rm~ergs~s^{-1}~kpc^{-2})$  & $0.62_{-0.11}^{+0.14}$ \\
$\beta$  & $0.397\pm0.009$ \\
$r_{\rm core}/r_{\rm 200}$ & 0.002 \\
$n_{\rm 0}/(f^{-1/2}\rm~cm^{-3})$  & $11.76_{-1.06}^{+1.30}$ \\
$P_{\rm 0}/(f^{-1/2}\rm~eV~cm^{-3})$ & $18.62_{-1.69}^{+2.05}$ \\
$t_{\rm cool,0}/(f^{1/2}\rm~Gyr)$ & $0.90_{-0.10}^{+0.08}$ \\
$N_{\rm p,0}/(f^{-1/2}\rm~10^{20}~cm^{-2})$ & $3.70_{-0.54}^{+0.68}$ \\
$L_{\rm X,r<0.1r_{200}}/(10^{40}\rm~ergs~s^{-1})$ & $0.75_{-0.13}^{+0.16}$ \\
$L_{\rm X,r<r_{200}}/(10^{40}\rm~ergs~s^{-1})$ & $3.11_{-0.56}^{+0.68}$ \\
$M_{\rm hot,r<r_{200}}/(f^{1/2}\rm~10^{11}~M_\odot)$ & $1.26_{-0.11}^{+0.14}$ \\
$E_{\rm hot,r<r_{200}}/(f^{1/2}\rm~10^{59}~ergs)$ & $4.52_{-0.41}^{+0.50}$ \\
$r_{\rm cool}/\rm kpc$  & $6.46_{-0.47}^{+0.53}$ \\
$\dot{M}_{\rm cool,r<r_{\rm cool}}/(\rm M_\odot~yr^{-1})$ & $0.012\pm0.002$ \\
$M_{\rm b,r<r_{200}}/(\rm 10^{11}~M_\odot)$ & $4.32\pm1.53$ \\
$f_{\rm b}$     & $(4.5\pm2.6)\%$ \\
$f_{\rm b,hot}$     & $(1.3\pm0.6)\%$ \\
$F_{\rm hot}$     & $(29.2_{-2.6}^{+3.2})\%$ \\
$F_{\rm b,detect}$     & $(27.2\pm15.6)\%$ \\
$F_{\rm b,missing}$     & $(72.8\pm15.6)\%$ \\
\hline
$M_{\rm hot,flat,r<r_{200}}/(f^{1/2}\rm~10^{11}~M_\odot)$ & $3.25_{-0.28}^{+0.34}$ \\
$E_{\rm hot,flat,r<r_{200}}/(f^{1/2}\rm~10^{59}~ergs)$ & $11.6_{-1.0}^{+1.2}$ \\
$M_{\rm b,flat,r<r_{200}}/(\rm 10^{11}~M_\odot)$ & $6.30_{-1.55}^{+1.56}$ \\
$f_{\rm b,flat}$     & $(6.5\pm3.3)\%$ \\
$f_{\rm b,hot,flat}$     & $(3.3_{-1.5}^{+1.6})\%$ \\
$F_{\rm hot,flat}$     & $(51.5_{-4.5}^{+5.4})\%$ \\
$F_{\rm b,detect,flat}$     & $(38.9_{-19.8}^{+19.9})\%$ \\
$F_{\rm b,missing,flat}$     & $(61.1_{-19.9}^{+19.8})\%$ \\
\hline
\end{tabular}\label{table:FiducialGalaxy}
\end{center}
\noindent 
$M_*$, $v_{\rm rot}$, $M_{\rm 200}$, and $r_{\rm 200}$ are the average parameters of the CGM-MASS galaxies. $kT$ and $Z$ are consistent with results from spectral analysis (Fig.~\ref{fig:CGMMASSSpec}). $I_{\rm 0}$, $\beta$, and $r_{\rm core}$ are the parameters of the radial intensity profile (Fig.~\ref{fig:rProfileStack}b). $n_{\rm 0}$, $P_{\rm 0}$, $t_{\rm cool, 0}$, $N_{\rm p,0}$ are the hydrogen number density, thermal pressure, radiative cooling timescale, and hydrogen column density of hot gas at the center of the galaxy ($r=0$), which, together with $\beta$ and $r_{\rm core}$, can be used to characterize the radial distribution of hot gas using the equations listed in \citep{Li17}. $L_{\rm X}$ is the extinction-corrected 0.5-2~keV luminosity of hot gas. $M_{\rm hot}$, $E_{\rm hot}$ are the total mass and thermal energy. $r_{\rm cool}$ is the cooling radius at which the radiative cooling timescale $t_{\rm cool}\sim10\rm~Gyr$. $\dot{M}_{\rm cool,r<r_{\rm cool}}$ is the radiative cooling rate calculated within $r_{\rm cool}$. $M_{\rm b}$ and $f_{\rm b}$ are the total baryon mass and baryon fraction including hot gas and stellar masses. $F_{\rm hot}$ is the fraction of baryon detected in hot phase defined as: $F_{\rm hot}=M_{\rm hot}/(M_{\rm hot}+M_*)$. $F_{\rm b,detect}$ and $F_{\rm b,missing}$ are the detected and missing fraction of baryons for the fiducial galaxy. We have assumed the volume filling factor of hot gas $f=1$ when calculating $f_{\rm b}$ and $F_{\rm hot}$. The parameters below the horizontal line with a label ``flat'' are estimated based on a flattened density profile with the best-fit $\beta$-function at $r<0.3r_{\rm 200}$ and a constant density at $r=(0.3-1.0)r_{\rm 200}$. 
\\
\end{table}

The cosmic baryon fraction, or the baryon-to-total (baryon+dark matter) mass ratio, is $(16.69\pm0.63)\%$ \citep{Komatsu09}. By extrapolating the X-ray intensity profile to $r_{\rm 200}$, the derived mass of the hot CGM accounts for $(7.8\pm3.6)\%$ of the expected baryons. Assuming the mass of other cooler gas phases are negligible for such massive quiescent galaxies (e.g., \citealt{Li16b}), the total (stellar+hot gas) baryons detected within $r_{\rm 200}$ accounts for $(27\pm16)\%$ of the expected baryons, implying that $\sim73\%$ of the baryons are still ``missing'' from the current survey of baryons in stars and the hot CGM. Compared to the $L^\star$ galaxy studied by the COS-Halos group, which has only $<6\%$ of the baryons stored in the extended hot CGM \citep{Werk14}, our fiducial galaxy has $\sim29\%$ of the baryons in hot phase, or the total mass of the extended hot halo is $\sim45\%$ of the stellar mass of the galaxy within $r_{\rm 200}$. 

As we have only directly detected the hot gas to $r\approx0.3r_{\rm 200}$ above the 1~$\sigma$ background uncertainty (Fig.~\ref{fig:rProfileStack}b), we also estimate the upper limit of the hot CGM mass by adopting the best-fit $\beta$-function at $r<0.3r_{\rm 200}$ and a constant density at $r=(0.3-1.0)r_{\rm 200}$. The estimated baryon mass and other related parameters are also listed in Table~\ref{table:FiducialGalaxy}. In particular, the firm upper limit of the hot baryon fraction obtained from this flattened profile is $f_{\rm b,hot,flat}=(3.3_{-1.5}^{+1.6})\%$, or $\approx20\%$ of the expected baryons. Therefore, at least $\sim60\%$ of the baryons are still ``missing''.

A key result of the stacking analysis is the constant slope of the radial soft X-ray intensity profile at $r\lesssim 0.5r_{\rm 200}$. Some models predict a flattened X-ray intensity profile at larger radii (e.g., \citealt{Faerman17}; Fig.~\ref{fig:rProfileStack}b), while many simulations produce declined profiles at $r\gtrsim(0.1-0.2)r_{\rm 200}$ (e.g., \citealt{Crain13}). The slope of the X-ray intensity profile could be modified by the combination of a few effects: (1) star formation and AGN feedback flattens the X-ray intensity profile in the inner region by preferentially removing low entropy gas from the hot halo; (2) the X-ray emission is proportional to both the metallicity and density, so the change of either of them may modify the slope of the X-ray intensity profile. The observed constant slope of the X-ray intensity profile is suggestive of both a weak impact of feedback in the recent past, and a constant metallicity or a flattened density profile (if the metallicity is declining) at larger radii.

\begin{figure*}
\begin{center}
\epsfig{figure=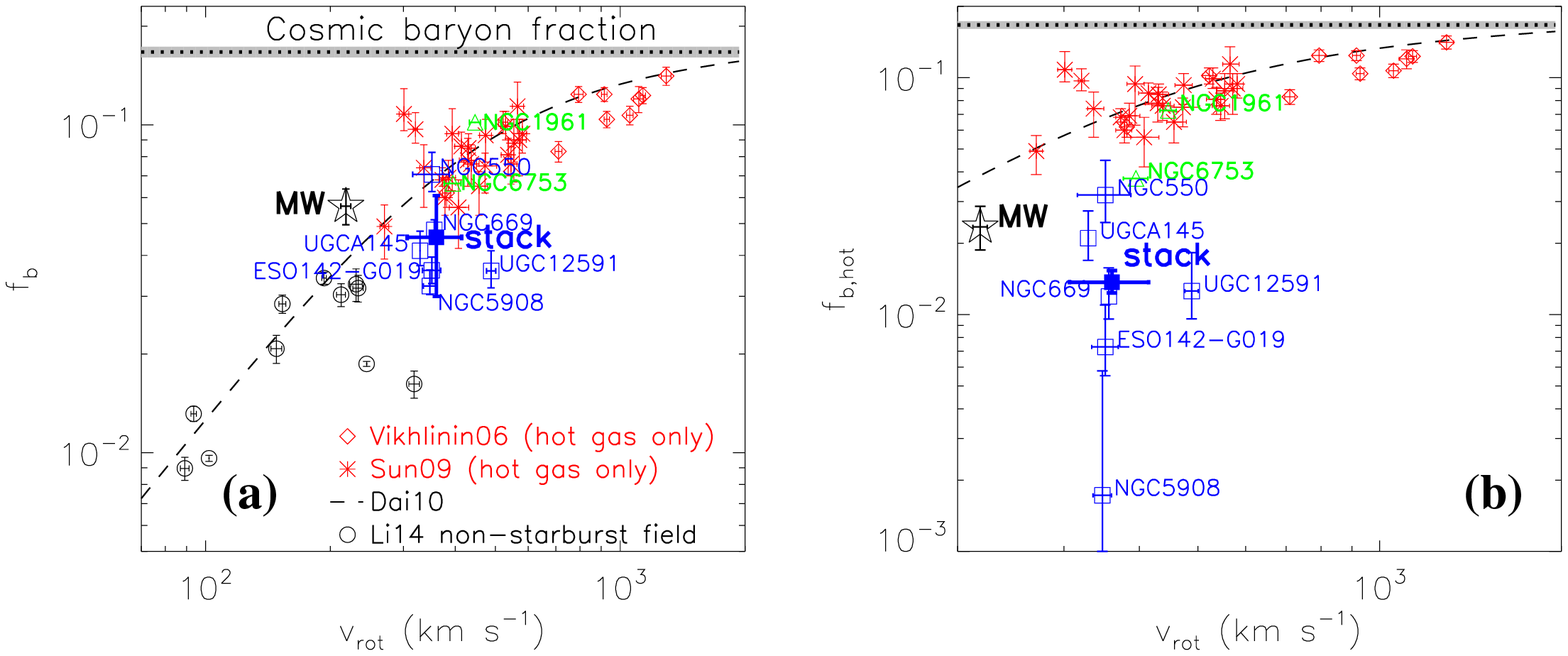,width=1.0\textwidth,angle=0, clip=}
\caption{\textbf{(a)} Baryon fraction ($f_{\rm b}$) v.s. rotation velocity ($v_{\rm rot}$). The cosmic baryon fraction (dotted line) with error (shaded area), a fitted relation from \citet{Dai10} (dashed), the MW \citep{Miller15} (black five-pointed star), samples of non-starburst field spirals \citep{Li14} (black circles), galaxy groups \citep{Sun09} (red stars), and galaxy clusters \citep{Vikhlinin06} (red diamonds) are plotted for comparison. For the non-starburst field spirals, only the stellar mass and the hot gas within a few tens of kpc are included in the baryon budget; an extended hot halo is not detected. For galaxy groups and clusters, only the hot gas component (no stellar component) is included, and we have simply assumed the rotation velocity equals to the velocity dispersion for a qualitative comparison. \textbf{(b)} Hot gas baryon fraction ($f_{\rm b,hot}$) v.s. $v_{\rm rot}$. We include only the extended hot gas component for a uniform comparison, with no stellar component and no low mass galaxies without a detected extended hot halo.}\label{fig:baryon}
\end{center}
\end{figure*}

We further compare the baryon budget of the fiducial galaxy and individual CGM-MASS galaxies to other galaxies \citep{Bogdan13,Miller15,Li13a,Li13b,Li14,Anderson16}, galaxy groups \citep{Sun09}, and galaxy clusters \citep{Vikhlinin06}. We see from Fig.~\ref{fig:baryon}a that the massive spiral galaxies (CGM-MASS, NGC~1961, NGC~6753) generally follow the same trend as lower mass galaxies and more massive groups and clusters of galaxies on the  $f_{\rm b}-v_{\rm rot}$ relation. The increase of $f_{\rm b}$ at higher $v_{\rm rot}$ indicates that more massive halos are closer to a ``closed box'' for baryonic matter, compared to $L^\star$ galaxies such as the MW. For massive spiral galaxies, the hot baryon fraction $f_{\rm b,hot}$ still remains significantly lower than inferred from X-ray observations of galaxy groups with similar $v_{\rm rot}$ (Fig.~\ref{fig:baryon}b), although recent SZ measurements of galaxy groups indicate a slightly lower hot gas content better matching our X-ray measurements of the hot CGM around massive spiral galaxies \citep{Lim17a}.

Explanations for the undetected baryons include: ejection from the halo by energetic feedback acting throughout the assembly of the galaxies (e.g., \citealt{McCarthy11}); remaining in a cooler phase that is difficult to detect (e.g., \citealt{Tumlinson11,Lim17b}); or having never been accreted onto the halo due to the heating of the gas by the collapse of large-scale structures prior to the assembly of the halos (e.g., \citealt{Mo05}). 
Most of the SZ measurements favor a large fraction of the gas being in a hot phase, either around individual galaxies, groups, or clusters (e.g., \citealt{Planck13,Lim17a}), or in the large scale cosmic web \citep{deGraaff17}. 
We have shown in a companion paper \citep{Li17} that the \emph{current} \emph{stellar} feedback rate in the CGM-MASS galaxies is unlikely energetic enough to expel a significant fraction of the baryons beyond the halo ($r_{\rm 200}$), but it is likely that the ejection took place primarily at earlier epochs or by occasional AGN episodes. The virial temperatures of the CGM-MASS galaxies are also sufficiently high that gas accreted onto the halo is heated gravitationally to an X-ray emitting temperature above the peak of the radiative cooling curve at $T\sim10^{5-6}\rm~K$, leaving little volume for cooler gas phases such as detected in $L^\star$ galaxies (e.g., \citealt{Tumlinson11,Werk14}) or claimed in more massive systems (e.g., \citealt{Lim17b}). The radiative cooling timescale of the halo gas is also too long for the gas condensation and precipitation to be important \citep{Li17}, which is often adopted to explain the detection of cool gas in massive elliptical galaxies (e.g., \citealt{Voit15}). A pre-virialization gravitational heating scenario is unlikely to be sufficient to prevent the collapse of gas, especially in high-mass halos (e.g., \citealt{Crain07}). Therefore, the significantly lower hot gas content of isolated massive spiral galaxies relative to galaxy groups with a similar gravitational potential indicates that the fraction of ``missing baryons'' at these mass scales is sensitive to the environment and/or the assembly history of the system. An actively assembling system such as merging galaxies or groups/clusters of galaxies often exhibits extremely strong feedback that can expel a majority of baryons from the halo at high redshift, leaving a quiescent descendant with a gas-deficient halo.

\bigskip
\noindent\textbf{\uppercase{acknowledgments}}
\smallskip\\
\noindent JTL and JNB acknowledge the financial support from NASA through the grants NNX15AM93G, NNX15AV24G, \emph{SOFIA} 05-0020, \emph{Chandra} 18610344, and \emph{XMM-Newton} 080073. QDW is supported by NASA via a subcontract of the grant NNX15AM93G. RAC is a Royal Society University Research Fellow.

\end{document}